\documentstyle[12pt,aps,prc,epsfig]{revtex}
 \tightenlines

\newpage
\begin{document}

\makeatletter
\def\@fnsymbol#1{\ifcase#1\or *\or \dagger\or 1\or 2\or
3\or 4\or 5 \or \ddagger\ddagger \or \mathchar "278 \mathchar
"278\or ***\or\dagger\dagger\dagger\or\ddagger\ddagger\ddagger\or
\mathchar "278 \mathchar "278 \mathchar "278
\else\@ctrerr\fi\relax} \makeatother

\centerline{ \bf Semiempirical Shell Model Tabulated Masses for
Translead Elements} \centerline{\bf
 with Magic Proton Number Z = 126 \footnote{This work has been
 submitted to Academic Press for possible publication in ADNDT.
  Copyright may be transferred without notice,
  after which this version may no longer be accessible.}}

\medskip

\centerline{\it S. LIRAN \footnote{Present address: Kashtan 3/3,
Haifa 34984, Israel} , A. MARINOV and N. ZELDES}

\centerline{\it The Racah Institute of Physics, The Hebrew
University of Jerusalem,}

\centerline{\it Jerusalem 91904, ISRAEL}

\begin{abstract}
We present two tables of calculated masses of translead nuclei,
for 351 nuclei with 94 $\leq$ N $\leq$ 126, 82 $\leq$ Z $\leq$ 100
and for 1969 nuclei with 126 $\leq$ N $\leq$184, 82 $\leq$ Z
$\leq$ 126. The tables are calculated from a semiempirical
shell-model mass equation based on Z = 126 as a proton magic
number which seems to be highly extrapolatable inside shell
regions. Useful separation and decay energies are given as well.
Some properties of the calculated masses and applications to
superheavy elements are indicated in the introduction.
\end{abstract}\


\vspace*{1.0cm} \centerline{CONTENTS}

\vspace*{0.5cm} INTRODUCTION

\hspace*{1.0cm} 1. Overview

\hspace*{1.0cm} 2. The Mass Equation

\hspace*{1.0cm} 3. Extrapolatability of the Mass Equation

\hspace*{1.0cm} 4. Two Weak Points

\hspace*{1.5cm} 4.1. Over-smoothness inside shell regions

\hspace*{1.5cm} 4.2. Discontinuities along shell region boundaries

\hspace*{1.0cm} 5. Illustrative Applications

 EXPLANATION OF TABLES

TABLE A: Atomic mass excesses and mass differences in keV for
nuclei in the shell region

\hspace*{1.9cm} with 82 $\leq$ N, Z $\leq$ 126.

TABLE B: Atomic mass excesses and mass differences in keV for
nuclei in the shell region

\hspace*{1.9cm} with 126 $\leq$ N $\leq$ 184, 82 $\leq$ Z $\leq$
126.

 \vspace*{1.0cm} \centerline{\bf{INTRODUCTION}}
\vspace*{0.3cm} {\bf 1. Overview} \vspace*{0.2cm}

The present mass tables are offered as a substitute and extension
for translead nuclei of the semiempirical shell-model mass
equation (SSME) table presented in ref. \cite{lz76}. They have
both been calculated in the same framework of the semiempirical
shell model \cite{ze96}, with the difference that here the major
proton valence shell beyond lead extends through Z = 126 rather
than Z = 114. This extends the range of applicability of the
equation from Z = 114 to Z = 126.

The need to go beyond Z = 114 arises from recent experimental
results on superheavy elements (SHE) \cite{nin99}, where the
nucleus $^{293}$118 was presumably formed and $\alpha$-decayed
sequentially down to $^{265}$Rf (Z = 104), with $\alpha$-decay
energies varying rather smoothly along the chain. If the results
are confirmed, and the decaying nuclei are formed in or near their
ground states (g.s.), then the smooth variation seems to preclude
the traditional macroscopic-microscopic \cite{mon94,smo97} Z = 114
as a major proton magic number in these nuclei, and suggests that
all of them belong to the same major proton valence shell
extending up to Z = 118 or beyond.

Moreover, the SSME \cite{lz76} which adopts Z = 114 as an upper
proton shell boundary and stops there becomes unsuitable for
extrapolation to higher Z-values already earlier, beyond Hs (Z =
108), as shown by its increasing deviations from the data when Z
increases \cite{nin99,lmz00b}.

Which is the major spherical proton magic number after lead?
During the early stages of developing the SSME \cite{li73}, when
it was adjusted separately in individual shell regions in the N -
Z plane, both Z = 114 and Z = 126 were considered possible
alternative candidates for the post lead proton magic number
\cite{kum89}. They were both tried as a shell boundary in each of
the two heaviest regions with Z $\geq$ 82 and respective N
boundaries 82 $\leq$ N $\leq$ 126 (called here region A) and 126
$\leq$ N $\leq$ 184 (called region B). The agreement with the data
was about the same for both choices, and the prevailing view in
the mid nineteen-seventies led to the choice of Z = 114 for the
SSME mass table.

When the Z = 118 results were obtained the SSME \cite{lz76} could
not reproduce them \cite{nin99} but the early Z = 126 results
agreed with them rather well \cite{lmz00a}. Phenomenological
studies of B(E2) systematics \cite{zam95} likewise indicate a
superior magicity of Z = 126 as compared to Z = 114, and the
plausibility of Z = 126 as a spherical proton magic number after
lead is indicated as well by the persistence of the Wigner term in
masses of heavy nuclei \cite{zel98}. The masses presented in the
present tables have been calculated on this assumption. For region
B they are the same as calculated in ref. \cite{li73}. For region
A a partial readjustment of the coefficients  was found necessary
 \cite{lmz01} (see sect. 3).

Recent self-consistent (SC) and relativistic mean field
calculations \cite{cwi96,ben99,kru00} variously predict proton
magicity for Z = 114, 120, 124 and 126, depending on the
interaction used. In this connection it is worthwhile emphasizing
that the rather suggestive agreement with the data obtained with Z
= 126 is not a proof of superior magicity of Z = 126 as compared
to Z = 120 or 124, because no comparative mass studies of this
kind were made.

The rest of the Introduction is organized as follows: The mass
equation is presented in sect. 2. Sect. 3 addresses its
extrapolatability and sect. 4 considers its smoothness and
continuity. Finally, sect. 5 briefly illustrates some applications
to SHE.\\

{\bf 2. The Mass Equation} \vspace*{0.2cm}

 In the SSME the total
nuclear energy in the g.s. is written \cite{lz76,ze96} as a sum of
pairing, deformation and Coulomb energies:

\begin{equation}
E\left( {N,Z} \right)=E_{pair}\left( {N,Z} \right)+E_{def}\left(
{N,Z} \right)+E_{Coul}\left( {N,Z} \right)\ . \label{eq1}
\end{equation}

The form of E$_{Coul}$, which describes the Coulomb energy of the
protons, is the same in all shell regions:

\begin{equation}
E_{Coul}\left( {N,Z} \right)=\left( {{{2Z_0} \over A}}
\right)^{1/3} [{\alpha ^C+\beta ^C\left( {Z-Z_0} \right)+\gamma
^C\left( {Z-Z_0} \right)^2}]\ . \label{eq2}
\end{equation}

The form of E$_{pair}$, which describes the energy of strongly
interacting nucleon pairs in a lowest seniority approximation, is
the same separately in all diagonal shell regions, where the major
valence shells are the same for neutrons and protons, and in all
non-diagonal regions, where they are different. Unlike in ref.
\cite{lz76}, with Z = 126 rather than 114 as an upper proton
boundary region A becomes a diagonal region with

\begin{eqnarray}
E_{pair}\left( {N,Z} \right) &=& \left( {{{A_0} \over A}}
\right)[\alpha +\beta \left( {A-A_0} \right)+\gamma \left( {A-A_0}
\right)^2+\varepsilon T\left( {T+1} \right) \nonumber \\
  &&+{{1-\left( {-1} \right)^A} \over 2}\Theta +{{1-\left( {-1}
\right)^{NZ}} \over 2}\kappa ]\ . \label{eq3}
\end{eqnarray}

\noindent{For the non-diagonal region B one has:}

\begin{eqnarray}
    E_{pair}(N,Z) &=& \left({{A_{0}}\over{A}}\right)[ \alpha +
\beta_{1} (N -
   N_{0}) + \beta_{2}(Z - Z_{0})  \quad  \nonumber \\
     &&+ \gamma_{1} (N - N_{0})^{2} + \gamma_{2}(Z - Z_{0})^{2}+
     \gamma_{3}(N-N_{0})(Z- Z_{0}) \quad \nonumber \\
     && +{{1 - (-1)^{N}}\over{2}}\Theta_{1} +{{1 -
     (-1)^{Z}}\over{2}}\Theta_{2} +{{1 - (-1)^{NZ}}\over{2}}\mu  ]
     \quad .
    \label{eq4}
\end{eqnarray}

The part E$_{def}$ describes additional negative energy due to
configuration interaction, largely  with pair breaking and
deformation. For region A it is given by \cite{li73,lmz01}

\begin{equation}
E_{def}\left( {N,Z} \right)=\left( {{{A_0} \over A}} \right)\left[
{\varphi _{11}\Phi _{11}\left( {N,Z} \right)+\psi _{20}\left[
{\Psi _{20}\left( {N,Z} \right)+\Psi _{20}\left( {Z,N} \right)}
\right]} \right]
    \label{eq5}
\end{equation}

with

\begin{equation}
\Phi _{11}\left( {N,Z} \right)=\left( {N-82} \right)\left( {126-N}
\right)\left( {Z-82} \right)\left( {126-Z} \right) \ ,
    \label{eq6}
\end{equation}

\begin{equation}
\Psi _{20}\left( {N,Z} \right)=\left( {N-82} \right)^2\left(
{126-N} \right)^2\left( {N-104} \right) .
    \label{eq7}
\end{equation}

\noindent{For region B it is given by \cite{li73,lmz00a}

\begin{equation}
    E_{def}(N,Z)=\left({{A_{0}}\over{A}}\right)\left[\varphi_{21}\Phi_{21}(N,Z)+
\varphi_{31}\Phi_{31}(N,Z)+\chi_{12}X_{12}(N,Z)\right]
    \label{eq8}
\end{equation}

with

\begin{equation}
    \Phi_{21}(N,Z) = (N - 126)^{2}(184 - N)^{2}(Z - 82)(126 - Z) \ ,
    \label{eq9}
\end{equation}
\begin{equation}
    \Phi_{31}(N,Z) = (N - 126)^{3}(184 - N)^{3}(Z - 82)(126 - Z) \ ,
    \label{eq10}
\end{equation}
\begin{equation}
    X_{12}(N,Z) = (N - 126)(184 - N)(N - 155)(Z - 82)^{2}(126 -
    Z)^{2}(Z - 104)\ .
    \label{eq11}
\end{equation}

 In eqs.~(2)-(5) and (8) $A=N+Z$ and $T = |T_{z}| =
{1 \over 2} |N - Z|$~\footnote{In the as yet unknown odd-odd $N =
Z$ translead nuclei the g.s. is expected to have $T = |T_{z}|+1$
and seniority zero, whereas eq.~(1) with $T = |T_{z}|$ gives the
energy of a low excited seniority two state \cite{ze96}.}. The
respective values of ($N_{0}, Z_{0}, A_{0}$) in regions A and B
are (82, 82, 164) and (126, 82, 208). The coefficients multiplying
the functions of N and Z are adjustable parameters determined by a
least squares adjustment to the data, separately for region B
\cite{li73} and for region A \cite{li73,lmz01}. Their  values are
given in table I. The atomic mass excesses $\Delta$M(N,Z) are
obtained by adding to the adjusted energies  E(N,Z) the sum of
nucleon mass excesses N$\Delta$M$_{n}$ + Z$\Delta$M$_{H}$.

In region B the equation has 15 adjustable parameters and it has
respective overall average and root-mean-square (rms) deviations
of 13 and 156 keV from the 267 presently known masses, and
corresponding -5 and 178 keV from the 231 know Q$_\alpha$ values.
In region A there are 11 adjustable parameters, and respective
overall average and rms deviations of 2 and 246 keV from the 150
known masses, and 2 and 99 keV from the 109 known Q$_\alpha$
values. More details are given in sect. 3.

Mass predictions calculated from the above equations for the
respective regions A and B are given in tables A and B. Useful
separation and decay energies connecting nuclei in the same region
are given as well. The tables include particle-stable nuclei and
proton-unstable ones a short distance beyond the even proton drip
line.\\
\begin{minipage}{1\textwidth} 
\renewcommand{\footnoterule}{\kern -3pt} 

\begin{table}
    \caption{Values of the coefficients of
    eq.~(1) determined by adjustment to the data.}
\vspace*{0.2cm}
    \begin{tabular}{crcccr}
    &&&&& \\
    \multicolumn{3}{c}{Region B \cite{li73}} &\multicolumn{3}{c}{Region
    A\cite{li73,lmz01}} \\
        Coefficient & Value (keV) &&& Coefficient & Value (keV)  \\
        \hline
        $\alpha$ & $-2.3859605 \times 10^{6}$  &&&$\alpha$ &
$-1.987628 \times 10^{6}$\\
        $\beta_{1}$ & $-1.496441 \times 10^{4}$ &&&$\beta$ &
$-2.4773664 \times 10^{4}$ \\
        $\beta_{2}$ & $-3.3866255 \times 10^{4}$ &&&$\gamma$ &
$-8.51085 \times 10^{1}$\\
        $\gamma_{1}$ & $3.022233 \times 10^{1}$ &&&$\varepsilon$ &
$4.585496 \times 10^{2}$ \\
        $\gamma_{2}$ &  $2.811903 \times 10^{1}$ &&& $\Theta$ &
$1.2183 \times 10^{3}$\\
        $\gamma_{3}$ &  $-3.6159266 \times 10^{2}$ &&& $\kappa$ &
$2.1937 \times 10^{3}$\\
        $\Theta_{1}$ & $8.16 \times 10^{2}$ &&& $\alpha^{C}$ &
$7.968418 \times 10^{5}$ \\
        $\Theta_{2}$ & $1.007 \times 10^{3}$ &&& $\beta^{C}$ &
$2.032906 \times 10^{4}$ \\
        $\mu$ & $-2.121 \times 10^{2}$ &&&$\gamma^{C}$ & $9.819137
\times 10^{1}$ \\
        $\alpha^{C}$ & $8.111517 \times 10^{5}$ &&&$\varphi_{11}$ &
$-4.794 \times 10^{-2}$  \\
        $\beta^{C}$ & $2.0282913 \times 10^{4}$ &&& $\psi_{20}$ &
$9.095 \times 10^{-4}$ \\
        $\gamma^{C}$ & $1.0930065 \times 10^{2}$ &&&&  \\

        $\varphi_{21}$ & $-9.87874 \times 10^{-5}$ &&&&  \\

        $\varphi_{31}$ & $3.13824 \times 10^{-8}$ &&&& \\

        $\chi_{12}$ & $-1.428529 \times 10^{-7}$ &&&&  \\
   \end{tabular}
\end{table}
\renewcommand{\footnoterule}{\kern-3pt \hrule width .4\columnwidth
\kern 2.6pt}            
\end{minipage}

{\bf 3. Extrapolatability of the Mass Equation}
 \vspace*{0.2cm}

 We discuss extrapolatability by considering the new data measured
after the 1973 adjustments \cite{li73}, like in refs.
\cite{lmz00a,lmz01,hau84,mol95,mnk97}. We do it separately for
each region, starting with region B.

The experimental data used in the adjustments in region B included
211 masses (ref. \cite{wag71} augmented by data from the
literature up to Spring 1973). Presently there are 267 known
masses (ref. \cite{auw95} (excluding values denoted
``systematics'' (\#)) and recent literature). They include 56 new
masses that were not used in the adjustments.

Fig. 1 shows the deviations from the data of the predictions of
eq. (1) for the 56 newer masses, plotted   as function of the
distance from the line of $\beta$-stability,  NFS = $N - Z -
\frac{0.4A^{2}}{A + 200}$ \cite{hau84}. Empty circles denote the
deviations of the $N = 126 - 128$ nuclei $^{216}$Ac, $^{218}$Pa,
$^{216}$Th, $^{217}$Pa, $^{219}$Pa, $^{219}$U and $^{218}$U, which
increase in this order and indicate increasing underbinding of
extrapolated $N \approx 126$ nuclei when Z increases away from the
data. This will be further considered in subsect. 4.2.

\begin{figure}
\begin{center}
\leavevmode \epsfxsize=10.5cm \epsfbox{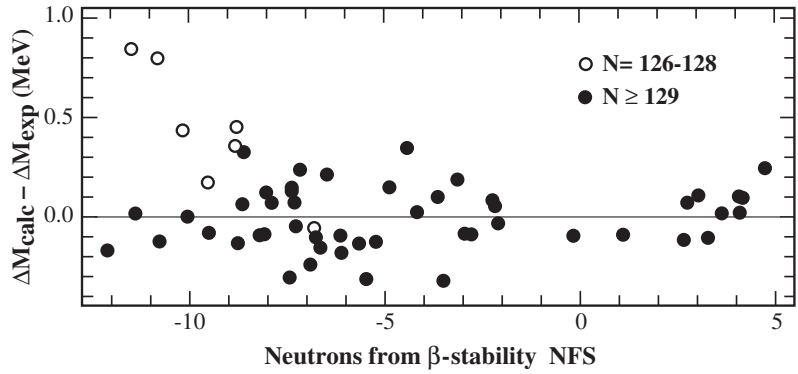}
\end{center}
\vspace*{-0.1cm}
 \caption{Deviations of the predicted masses
    (eq.~(1)) from the data for the 56 new masses in region B
    measured after the original adjustments were made. The deviations
    are plotted  as function of the variable NFS = N - Z
    - 0.4A$^{2}/(A+200)$ [16]. Taken from ref. [9].}
\end{figure}
\vspace{-0.2cm}
 The deviations of the remaining 49 nuclei with $N
\geq 129$, which do not follow the $N = 126$ boundary but extend
into the interior of the shell region, are marked by full circles.
They are about equally positive and negative, have similar
magnitudes, and do not seem to be correlated with NFS.

Table II, patterned after similar more elaborate ones
\cite{mol95,mnk97}, shows $\delta_{av}$ and $\delta_{rms}$, the
respective average and rms deviations of eq.~(1) from the data,
for $\Delta M, S_{n}, S_{p}, Q_{\beta^{-}}$ and $Q_{\alpha}$. The
deviations are shown separately for the older data that were used
in the 1973 adjustments and for the newer data. The last column
shows the error ratios $\delta_{rms}^{new}:\delta^{old}_{rms}$.\\
\begin{minipage}{1\textwidth} 
\renewcommand{\footnoterule}{\kern -3pt} 

\begin{table}
   \caption{Numbers of data N, average deviations $\delta_{av}$, and
    rms deviations $\delta_{rms}$, for eq.~(1) with the coefficients
    of table I for region B. The numbers in brackets are obtained when nuclei with
    $N = 126-128$ are excluded. The last column  shows the  ratios
    $\delta^{new}_{rms}:\delta^{old}_{rms}$. Taken from ref. [9].}
\vspace*{0.2cm}
    \begin{tabular}{lcccccccc}
    & \multicolumn{3}{c}{Original nuclei (1973)} & &
    \multicolumn{3}{c}{New nuclei (1973-1999)}  \\
         &  & $\delta_{av}$ & $\delta_{rms}$ & & & $\delta_{av}$ &
         $\delta_{rms}$ & Error  \\
        Data & N & (keV) & (keV) & & N & (keV) & (keV) & ratio \\
        \hline
        $\Delta M$ & 211 & ~~2 & 126 & & 56 (49) & 53 ($-1$) & 236
(155) &
        1.87 (1.23) \\
        $S_{n}$ & 169 & ~~1 & 117 &  & 45 (38) & 12 $(-2)$ & 171
(145) &
        1.46 (1.24) \\
        $S_{p}$ & 162 & $-4$ & 121 &  & 52 (44) & $-17$ (15) & 184
(148) &
        1.52 (1.22) \\
        $Q_{\beta^{-}}$ & 146 & $-7$ & 158 &  & 51 (44) & $-19$
(14) & 209
        (169) & 1.32 (1.07) \\
        $Q_{\alpha}$ & 174 & $-6$ & 162 &  & 57 (55) & $-3~(-8)$ & 220
        (220) & 1.36 (1.36) \\
    \end{tabular}
\end{table}
\renewcommand{\footnoterule}{\kern-3pt \hrule width .4\columnwidth
\kern 2.6pt}            
\end{minipage}

For the old data the magnitudes of  $\delta_{av}$ are single keVs,
and those of  $\delta_{rms}$  are in the range 110$-$170 keV. For
the new data they are larger, with respective highest values of 53
and 236 keV for $\Delta M$ and smaller values for $S_{n}, S_{p},
Q_{\beta^{-}}$ and $Q_{\alpha}$.

The table shows as well in brackets the corresponding deviations
for the 49 $N \geq 129$ nuclei extending into the interior of the
shell region, where SHE are presently searched for. Except for
$Q_{\alpha}$ they are smaller than the unbracketed deviations.

The deviations shown in table II are smaller than the
corresponding ones for several current mass models. This is
presumably mainly due to the inclusion in eq.~(1) of the
particle-hole(p-h)-symmetric configuration interaction terms
$E_{def}$ (eq.~(8)) \cite{lmz00a}. Configuration interaction is
largely missing in macroscopic-microscopic Strutinsky type and in
SC mean field calculations, where the included T = 1, J = 0
pairing correlations seem not to be enough. Based on the above
analysis we have recently \cite{lmz00a} proposed the use of the
masses given in table B as a substitute for the SSME \cite{lz76}
in the interior of region B.

The situation in region A is less simple. The experimental data
used in the 1973 adjustment included 29 masses and 62 $Q_{\alpha}$
values connecting unknown masses (ref. \cite{wag71} augmented by
data from the literature up to Spring 1973). Presently there are
150 known  masses  and 3 $Q_{\alpha}$ values connecting unknown
masses (refs. \cite{auw95} (excluding values denoted
``systematics'' (\#)) and \cite{radon} and recent literature).
There are 121 new masses that were not used in the adjustments.

Comparing the deviations of the predicted \cite{li73} 121 new
masses to those of the 29 original ones, one observes \cite{lmz01}
that the new deviations are as a rule considerably larger and
almost all negative, with respective average and rms values of
-807 and 1008 keV, as compared to -29 and 146 keV for the 29
original masses. For the 31 new Q$_\alpha$ values, though, the
deviations have perhaps even very slightly improved, becoming
respectively 40 and 89 keV as compared to 5 and 103 keV before.

Closer scrutiny \cite{lmz01} shows that the worse fit to the new
masses is mainly due to inadequate adjusted values of the
coefficients $\alpha$, $\varepsilon$, $\Theta$ and $\kappa$. These
coefficients largely cancel in Q$_\alpha$ and they were determined
essentially by the 29 original masses nearer to $\beta$-stability,
where the values of $\alpha$, $\Theta$ and $\kappa$ are smaller
(see ref. \cite{jhj84} for  $\Theta$) and that of $\varepsilon$ is
larger than for the 121 new nuclei nearer to the proton drip line.
Consequently a new least-squares adjustment of eq. (1) to all the
150 presently known masses was made \cite{lmz01}, with only four
adjustable parameters $\alpha$, $\varepsilon$, $\Theta$ and
$\kappa$ while the other seven parameters were held fixed on their
old adjusted values \cite{li73}. It was found that the new
adjusted coefficients have shifted in the expected directions, and
the resulting equation retains the high agreement with the data of
the old Q$_\alpha$ predictions, while at the same time the quality
of its agreement with the mass data has been largely improved. The
resulting set of coefficients for region A is the one given in
table I.

 Fig. 2 shows the deviations of the new predicted mass
values for all the 150 known masses. For ease of comparison empty
circles denote the deviations of the 29 originally adjusted masses
and full circles mark the deviations of the 121 new ones. As
already mentioned the deviations of the latter are considerably
smaller than for the original predictions. The deviations of the
29 older data have worsened, though.

\begin{figure}[h]
\begin{center}
\leavevmode \epsfxsize=10.5cm \epsfbox{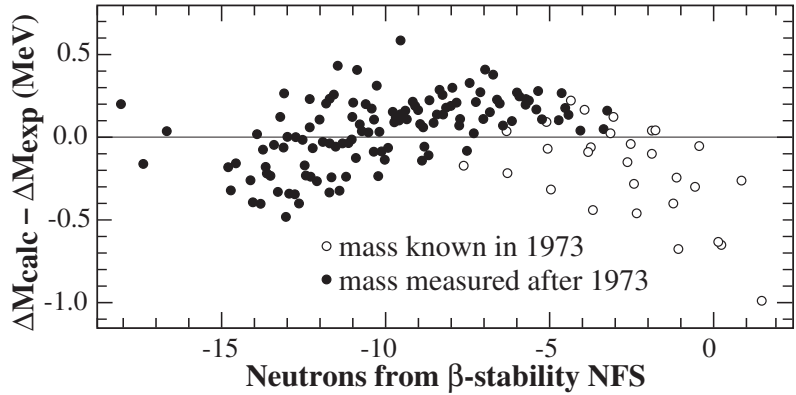}
\end{center} \caption{Deviations of the mass predictions
(eq.~(1)) from the data for
    all the 150 presently known masses in region A. The deviations are plotted as function of
the variable NFS = N - Z - 0.4A$^{2}$/(A+200) like in fig. 1.
Taken from ref. [12].}
\end{figure}

Table III shows the resulting $\delta$$_{av}$ and $\delta$$_{rms}$
values of the predicted deviations. They are shown separately for
the old 1973 data nearer to $\beta$-stability and for the new data
extending into the interior of region A towards the proton drip
line.  For S$_{n}$, S$_{p}$ and Q$_\alpha$ the new deviations,
shown in the last two columns of the table, are similar to those
of the original nuclei obtained in ref. \cite{li73}. For $\Delta$M
and $Q_{\beta^{-}}$ the $\delta$$_{rms}$ are respectively 1.5 and
1.1 times larger \cite{lmz01}.\\
\begin{minipage}{1\textwidth} 
\renewcommand{\footnoterule}{\kern -3pt} 

\begin{table}
    \caption{Numbers of data N, average deviations $\delta_{av}$, and
    rms deviations $\delta_{rms}$, for eq.~(1) with the new values of the
    coefficients $\alpha, \varepsilon, \Theta, \kappa$ and the old values
    of the other seven coefficients from Table I for region A. Taken from ref. [12].}

    \vspace*{0.2cm}
    \begin{tabular}{lccccccc}
    & \multicolumn{3}{c}{Original nuclei (1973)} & &
    \multicolumn{3}{c}{New nuclei (1973-2000)}  \\
         &  & $\delta_{av}$ & $\delta_{rms}$ & & & $\delta_{av}$ &
         $\delta_{rms}$  \\
        Data & N & (keV) & (keV) & & N & (keV) & (keV) \\
        \hline
        $\Delta M$ & 29 & $-193$ & 344 & & 121 & $48$ & 216  \\
        $S_{n}$ & 18 & 158 & 416 &  & 120 & $-10$ & 205  \\
        $S_{p}$ & 22 & $-144$ & 202 &  & 104 & 18 & 184  \\
        $Q_{\beta^{-}}$ & 15 & $-257$ & 475 &  & 101 & 15 & 277  \\
        $Q_{\alpha}$ & 78 & $-5$ & 104 &  & 31 & 18 & 85  \\
    \end{tabular}
\end{table}
\renewcommand{\footnoterule}{\kern-3pt \hrule width .4\columnwidth
\kern 2.6pt}            
\end{minipage}

Like in table II, the deviations shown in table III are smaller
than the corresponding ones for several current mass models. Based
on the above analysis we have recently \cite{lmz01} proposed the
use of the masses given in table A as a substitute for the SSME
\cite{lz76} in the interior of region A, particularly for the
extrapolatable-proven Q$_\alpha$ values.

\vspace{0.2cm}
 {\bf 4. Two Weak Points}

\hspace*{0.5cm} {\bf 4.1. Over-smoothness inside shell regions}

\vspace*{0.2cm}  Inside a shell region eq.~(1) describes a mass
surface which is smoother than the empirical surface and is
inadequate for describing fine structure effects
\cite{lz76,lmz00a}.

This is illustrated in fig. 3 showing Q$_\alpha$ systematics for
the heaviest N $\geq$ 140 even-Z nuclei from Pu through Z = 110
\cite{auw95}. Respective full and empty circles denote
experimental values and values estimated from systematics. The
small circles connected by thin lines show the predictions of
eq.~(1).

\begin{figure}[h]
\begin{center}
\leavevmode \epsfxsize=10.5cm \epsfbox{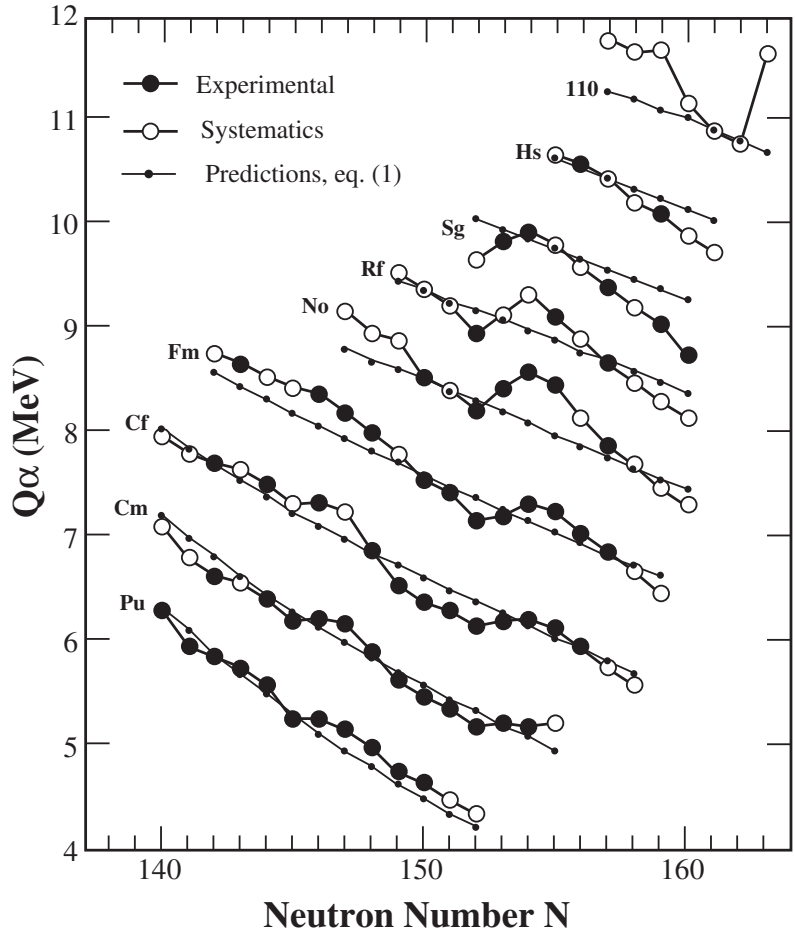}
\end{center} \caption{Q$_\alpha$ systematics of even-Z elements from Pu
through Z = 110 for N $\geq$ 140. Data taken from ref. [20] and
predictions from refs. [7,9].}
\end{figure}

As a rule the experimental isotopic lines show similar negative
trends when N increases, and they shift upwards rather uniformly
when Z increases. This regular pattern breaks down for nuclei in
the vicinity of the deformed doubly submagic nuclei (N$_{0}$,
Z$_{0}$) $^{252}$Fm (N$_{0}$ = 152, Z$_{0}$ = 100) and (presumably
even more so) $^{270}$Hs (N$_{0}$ = 162, Z$_{0}$ = 108)
\cite{mon94,smo97,laz96}, where the trend of isotopic lines
between N = N$_{0}$ and N = N$_{0}$ + 2 is positive, and the
vertical distance between isotopic lines with Z = Z$_{0}$ and Z =
Z$_{0}$ + 2 for N $\approx$ N$_{0}$ is larger than for other Z
values.

None of these submagic number effects is shown by the predicted
thin lines systematics.

 Another illustration of over-smoothness of eq.~(1) as compared to
 the data is seen in fig. 4. Additional examples are documented in
 ref. \cite{lz76}.

 In the SSME non-smooth abrupt local changes associated with
 subshell and deformation effects are assumed to have been
 smoothed out by configuration interaction, eqs.~(5) and (8), and
 the mass equation describes a smooth surface representing their
 average. The deviations from the average are mostly small,
 though, with an overall rms as given in table II.

\vspace*{0.2cm}
 \hspace*{0.5cm} {\bf 4.2. Discontinuities along shell region
 boundaries}

\vspace*{0.2cm}
 Because the mass surface was adjusted in regions A and B
 separately it has different upwards curvings in the two regions,
 resulting in an unphysical  discontinuity along the boundary line N = 126,
 which increases monotonically away from the data
 \cite{li73,lmz00a,ze67,com70}. This discontinuity is as a rule a
 few hundreds keV for the experimentally known N = 126 nuclides
 between Pb and U, and it reaches 1.3 MeV in $^{227}$Md, which is
 the heaviest N = 126 isotone included in the tables.\footnote{In
 refs. \cite{lz76} and \cite{com70} such unphysical discontinuities along shell
 region boundaries were avoided by adjusting to the data in the
 two regions simultaneously, with continuity requirements along
 the boundary imposed as additional constraints.}

 On the other hand, consistency with the shell model requires the
 occurrence of discontinuous drops of predicted nucleon separation
 energies when crossing a corresponding shell boundary towards
 heavier nuclei. In the present case the expected drops of two-neutron separation
 energies when crossing N = 126 into region B are observed for all
 the elements included in the tables.

 \vspace*{0.2cm}
 {\bf 5. Illustrative Applications}

 \vspace*{0.2cm}
The mass tables are intended to be used as a predictive tool in
the interiors of regions A and B. We briefly apply them to four
recent SHE experiments \cite{nin99,oga99,wilk00,oga01}.

In the first experiment \cite{nin99} cold fusion of $^{208}$Pb
target nuclei and 449 MeV bombarding $^{86}$Kr ions was studied.
Three observed seven-members $\alpha$-decay chains are consistent
with the formation of $^{293}$118 and its sequential decay down to
$^{265}$Rf (Z = 104).

The compound nucleus (CN) formed in the reaction is $^{294}$118.
According to the Pb and Kr masses \cite{auw95} and the predicted
mass of the CN from table B it is formed at an excitation energy
(E$^{x}$) of about 12 MeV, which allows it to emit one neutron,
leaving the evaporation residue (EVR) $^{293}$118 at E$^{x}$
$\leq$ 4 MeV. Both isotopes might be considered possible parents
of the $\alpha$-decaying chain.

For the assigned parent $^{293}$118, assuming that the decays
proceed through or near the g.s., the respective average and rms
deviations of the predicted Q$_\alpha$ values from the observed
ones are $\delta_{av}$ = -197 keV and $\delta_{rms}$ = 308 keV.
The $\delta_{rms}$ value is consistent with table II, but the
$\delta_{av}$ is too negative.The largest deviation of -735 keV
for $^{293}$118 might also be too negative.

Fig. 4 shows the measured and predicted Q$_\alpha$ values. One
observes the over-smoothness mentioned in subsect. 4.1, and the
large negative deviation of $^{293}$118. The authors of ref.
\cite{cnh99} mention the possibility that the observed transitions
may occur between structurally similar low lying
[611]$\frac{1}{2}$$^{+}$ Nilsson levels which in some of the
nuclei are not g.s. Using their calculated levels this would
reduce the $^{293}$118 deviation to -575 keV.

For the CN $^{294}$118 considered as a  parent the deviations are
larger: $\delta_{av}$ = -366 keV, $\delta_{rms}$ = 464 keV and
$\delta$($^{294}$118) = -1011 keV. This might perhaps lend some
additional support to the authors' assignment of $^{293}$118 as
the parent. The authors of ref. \cite{cnh99} likewise obtained
better agreement of their SC calculations with the $^{293}$118
scenario.

\begin{figure}
\begin{center}
\leavevmode \epsfxsize=9.3cm \epsfbox{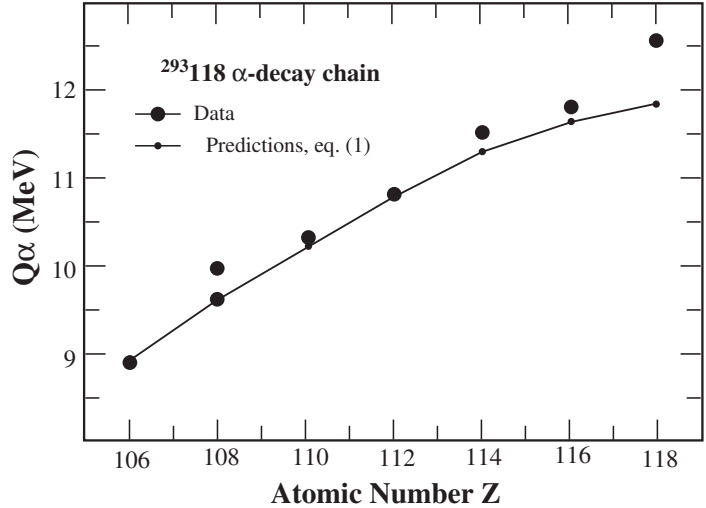}
\end{center}
\vspace*{-0.2cm}
 \caption{Experimental [3] and predicted [7,9] Q$_\alpha$ values of the
 $^{293}$118 decay chain.}
\end{figure}

The relations of the results shown in fig. 4 to several current
mass models are considered in ref. \cite{lmz00a}.

The CN and EVR parents, their corresponding formation channels,
estimated\footnote{The estimated values of E$^{x}$ in tables IV -
VII are obtained from the kinematics of the reactions  assuming
that the evaporated neutrons have zero kinetic energy and the
evaporated charged particles (p and $\alpha$) have a kinetic
energy which is equal to their potential energy at the top of the
Coulomb barrier. Higher kinetic energies of the evaporated
particles would reduce the estimates given in the tables.} values
of their excitation energies when formed, and the deviations
$\delta_{av}$ and $\delta_{rms}$ from the data of the
corresponding predicted Q$_{\alpha}$ values assuming that the
decays go through or near the g.s., are given in table IV.\\
\begin{minipage}{1\textwidth} 
\renewcommand{\footnoterule}{\kern -3pt} 

\begin{table}

 \caption{Conceivable parents of the $\alpha$-decay chains
[3] with their formation channels, deduced upper values (see
footnote 3) of the excitation energies E$^{x}$ of the radiative
capture and the
 evaporated residue nuclei, and the deviations $\delta_{av}$ and $\delta_{rms}$ from the
data of the corresponding  predicted Q$_\alpha$ values.}
\vspace*{0.2cm}
\begin{tabular}{lccll}
Parent & Channel & E$^{x}$ & $\delta_{av}$ & $\delta_{rms}$\\
  &  & (MeV) & (MeV) & (MeV)\\
\hline

 $^{294}$118 & CN & 12& -0.366 & 0.464\\

  $^{293}$118 & 1n & 4 & -0.197 & 0.308\\
\end{tabular}
\end{table}
\renewcommand{\footnoterule}{\kern-3pt \hrule width .4\columnwidth
\kern 2.6pt}            
\end{minipage}

The three other experiments were performed at higher excitation
energies of the CN. In the second experiment \cite{oga99} a
$^{244}$Pu target was bombarded by 236 MeV $^{48}$Ca ions. An
observed three-members $\alpha$-decay chain is considered a good
candidate for originating from the parent $^{289}$114 and its
sequential decay down to $^{277}$Hs (Z = 108).

The CN formed in the reaction is $^{292}$114 at E$^{x}$ $\approx$
27 MeV (ref. \cite{auw95} and table B). At this higher energy more
channels for particle emission might be open than in the Z = 118
cold fusion experiment, including up to 3n and also p or $\alpha$
emission. Four conceivable EVR parents, their formation channels,
estimated excitation energies at which they were formed, and the
deviations $\delta_{av}$ and $\delta_{rms}$ from the data of the
corresponding predicted Q$_\alpha$ values assuming that the decays
go through or near the g.s., are given in table V.\\
\begin{minipage}{1\textwidth} 
\renewcommand{\footnoterule}{\kern -3pt} 
\begin{table}
\caption{Conceivable EVR parents of the $\alpha$-decay chain [26]
with their formation channels, their estimated  values of E$^{x}$,
and the deviations $\delta_{av}$ and $\delta_{rms}$ from the data
of the corresponding predicted Q$_\alpha$ values.}
 \vspace*{0.2cm}
\begin{tabular}{lccll}
EVR & Evaporation & Estimated$^{a}$  E$^{x}$ & $\delta_{av}$ &
$\delta_{rms}$\\

 parent & channel & (MeV) & (MeV) & (MeV)\\
\hline

  $^{290}$114 & 2n & 16 & 0.643 & 0.720\\

  $^{289}$114 & 3n & 9 & 0.847 & 0.905\\

  $^{291}$113 & p & 8 & -0.241 & 0.414\\

   $^{288}$112 & $\alpha$ & 9 & -0.181 & 0.363\\
\end{tabular}
\vspace*{0.2cm}
 $^{a}$ See footnote 3.
\end{table}
\renewcommand{\footnoterule}{\kern-3pt \hrule width .4\columnwidth
\kern 2.6pt}            
\end{minipage}

In the third experiment \cite{wilk00} a $^{249}$Bk target was
bombarded by 117 MeV and 123 MeV $^{22}$Ne ions. We address the
five observed two-members $\alpha$-decay chains which are assigned
to the parent $^{267}$Bh (Z = 107).

The CN formed in the reaction is $^{271}$Bh at respective  E$^{x}$
values of about 43 and 48 MeV (ref. \cite{auw95} and table B),
which is higher than in the Z = 114 experiment, and presumably has
more allowed evaporation channels. Table VI, arranged like table
V, gives details for four conceivable EVR parents.\footnote{Other
EVRs, formed by $\alpha$2n and $\alpha$3n emissions, lead to
results which seem to be in conflict with known decay
characteristics of the nuclei involved.}\\
\begin{minipage}{1\textwidth} 
\renewcommand{\footnoterule}{\kern -3pt} 

\begin{table}

 \caption{Conceivable EVR parents of the two-members $\alpha$-decay
 chains
[27] with their formation channels, their estimated values of
E$^{x}$, and the deviations $\delta_{av}$ and $\delta_{rms}$ from
the data of the corresponding predicted Q$_\alpha$ values.}
 \vspace*{0.2cm}

\begin{tabular}{lccll}
EVR & Evaporation & Estimated$^{a,b}$  E$^{x}$  & $\delta_{av}$ &
$\delta_{rms}$\\

 parent & channel &(MeV) & (MeV) & (MeV)\\
\hline
 $^{267}$Bh & 4n & 18, 23 & 0.616 & 0.629\\

  $^{266}$Bh & 5n & 11, 16 & 0.712 & 0.723\\

  $^{268}$Sg & p2n & 14, 19 & -0.035 & 0.134\\

   $^{267}$Sg & p3n & 7, 12 & 0.066 & 0.146\\
\end{tabular}
\vspace*{0.2cm}  $^{a}$ See footnote 3.\\
 $^{b}$ The two estimated values correspond to the
two bombarding energies.
\end{table}
\renewcommand{\footnoterule}{\kern-3pt \hrule width .4\columnwidth
\kern 2.6pt}            
\end{minipage}

The first two lines in tables V and VI correspond to  xn emissions
including the EVR parents assigned by the authors. Their
deviations considerably exceed the values expected from table II
for g.s. transitions. If the above assignments are confirmed, the
large deviations might indicate that the decay chains do not
proceed through levels in the vicinity of the g.s.

The two subsequent lines in the tables correspond to conceivable
EVR parents formed by p, $\alpha$ and pxn emissions. Their
deviations are smaller, with $\delta_{rms}$ values consistent with
table II. This might lend some support to scenarios with pxn and
$\alpha$xn emissions, in addition to the pure xn evaporations
commonly considered.

Finally, in the fourth experiment \cite{oga01} a $^{248}$Cm target
was bombarded by 240 MeV $^{48}$Ca ions. An observed three-members
$\alpha$-decay chain is assigned to the nuclide $^{292}$116 and
its sequential decay down to $^{280}$110.

The CN formed in the reaction is $^{296}$116 at E$^{x}$ $\approx$
27 MeV (ref. \cite{auw95} and table B). Table VII gives details
for four conceivable EVR parents in addition to the assigned
parent $^{292}$116.\\
\begin{minipage}{1\textwidth} 
\renewcommand{\footnoterule}{\kern -3pt} 

\begin{table}

 \caption{Conceivable EVR parents of the $\alpha$-decay chain
[28] with their formation channels, their estimated  values of
E$^{x}$,  and the deviations $\delta_{av}$ and $\delta_{rms}$ from
the data of the corresponding predicted Q$_\alpha$ values.}
\vspace*{0.2cm}
\begin{tabular}{lccll}
EVR & Evaporation & Estimated$^{a}$  E$^{x}$  & $\delta_{av}$ &
$\delta_{rms}$\\

 parent & channel & (MeV) & (MeV) & (MeV)\\
\hline

  $^{294}$116 & 2n & 14 & 0.177 & 0.466\\

  $^{293}$116 & 3n & 7 & 0.423 & 0.577\\

  $^{292}$116 & 4n & 2 & 0.669 & 0.758\\

   $^{295}$115 & p & 7 & -0.767 & 0.905\\

      $^{292}$114 & $\alpha$ & 8 & -0.590 & 0.699\\
\end{tabular}
\vspace*{0.2cm}
 $^{a}$ See footnote 3.
\end{table}
\renewcommand{\footnoterule}{\kern-3pt \hrule width .4\columnwidth
\kern 2.6pt}            
\end{minipage}

The smallest deviations in increasing order occur for the
respective EVR parents $^{294}$116 and $^{293}$116 formed by 2n
and 3n emissions. Their $\delta_{rms}$ values are consistent with
table II. This might lend some support to a 2n (and possibly also
3n) scenario as compared to the 4n, p and $\alpha$ evaporation
channels.

The last two members of the $\alpha$-decay chain seen in this
experiment agree with the two-members $\alpha$-decay chains
observed before in a Z = 114 experiment \cite{oga01,oga00}. If
they are the same, a formation of the present Z = 116 parent by 2n
(or 3n) emission would imply the same formation channel(s) for the
Z = 114 EVR parent in ref. \cite{oga00}, rather than the assigned
4n channel.\\

We thank Yuri Lobanov and Yuri Oganessian for prepublication
results of ref. \cite{oga01} and a preprint of ref. \cite{oga00}
and for useful correspondence.

\newpage
\centerline{\bf{EXPLANATION OF TABLES}} \vspace{0.5cm}

Z \hspace{2.05cm}  Proton number.

N \hspace{2.0cm}    Neutron number.

A \hspace{2.0cm} Mass number: N + Z.

$\Delta$M(N,Z) \hspace{0.65cm} Atomic mass excess: E(N,Z) +
8071.323N + 7288.969Z keV, where E(N,Z)\\ \hspace*{3.1cm} is the
adjusted g.s. energy, eq. (1).

S$_{n}$(N,Z) \hspace{0.9cm} Neutron separation energy:
$\Delta$M(N-1,Z) - $\Delta$M(N,Z) + 8071.323 keV.

S$_{p}$(N,Z) \hspace{0.9cm} Proton separation energy:
$\Delta$M(N,Z-1) - $\Delta$M(N,Z) + 7288.969 keV.

Q$_\alpha$(N,Z) \hspace{0.8cm} Q-alpha value: $\Delta$M(N,Z) -
$\Delta$M(N-2,Z-2) - 2424.911 keV.

Q$_\beta$(N,Z) \hspace{0.8cm} Q-beta value: $\Delta$M(N,Z) -
$\Delta$M(N-1,Z+1).

S$_{2n}$(N,Z) \hspace{0.75cm} Two-neutron separation energy:
$\Delta$M(N-2,Z) - $\Delta$M(N,Z) + 16142.646 keV.

S$_{2p}$(N,Z) \hspace{0.75cm} Two-proton separation energy:
$\Delta$M(N,Z-2) - $\Delta$M(N,Z) + 14577.938 keV. \vspace{0.5cm}
\newpage

\newpage
\begin{table}[h]
 TABLE A. {\small Atomic mass excesses and mass differences in keV for
nuclei in the shell region\\ 82 $\leq$ N,Z $\leq$ 126, calculated
from eq.~(1) with the coefficients of region A in table~I.}
\vspace*{0.2cm}
\begin{tabular}{llllllllll}
Z & N & A & $\Delta$M & S$_n$ & S$_p$ & Q$_\alpha$ & Q$_\beta$ &
S$_{2n}$ & S$_{2p}$
\\
\hline\\[-10pt]
 82           & 94          & 176         & 10857       & 12871       &          &           &-15377       & 23676       & \\
 82           & 95          & 177         & 8541        & 10387       &          &           &-12831       & 23259       & \\
 82           & 96          & 178         & 4177        & 12435       &          &           &-14360       & 22822       & \\
 82           & 97          & 179         & 2263        & 9985        &          &           &-11857       & 22420       & \\
 82           & 98          & 180         &-1682        & 12017       &          &           &-13385       & 22002       & \\
 82           & 99          & 181         &-3215        & 9603        &          &           &-10927       & 21621       & \\
\end{tabular}
\begin{tabular}{llllllllll}
 88           & 102         & 190         & 38254       & 12664       &-858         & 9270        &-15272       & 23416       &-3433 \\
 88           & 103         & 191         & 35959       & 10366       &-835         & 9161        &-12948       & 23031       &-2955 \\
 88           & 104         & 192         & 31758       & 12271       &-408         & 9090        &-14403       & 22638       &-2508 \\
 88           & 105         & 193         & 29819       & 10010       &-391         & 8968        &-12122       & 22282       &-2047 \\
 88           & 106         & 194         & 25982       & 11908       & 25          & 8885        &-13579       & 21919       &-1617 \\
 88           & 107         & 195         & 24370       & 9682        & 37          & 8752        &-11339       & 21591       &-1173 \\
 88           & 108         & 196         & 20868       & 11573       & 444         & 8656        &-12799       & 21255       &-758 \\
 88           & 109         & 197         & 19558       & 9381        & 449         & 8514        &-10598       & 20954       &-331 \\
\end{tabular}
\renewcommand{\footnoterule}{\kern-3pt \hrule width .4\columnwidth
\kern 2.6pt}            
\end{table}
\newpage
\begin{table}[h]
 TABLE B. {\small Atomic mass excesses and mass differences in keV
for nuclei in the shell region\\ 126 $\leq$ N $\leq$ 184, 82
$\leq$ Z $\leq$ 126, calculated from eq. (1) with the coefficients
of region B in table I.}
\vspace*{0.2cm}
\begin{tabular}{llllllllll}
Z & N & A & $\Delta$M & S$_n$ & S$_p$ & Q$_\alpha$ & Q$_\beta$ &
S$_{2n}$ & S$_{2p}$
\\
\hline\\[-10pt]
 82           & 126         & 208         &-21909       &           &           &            &            &            &  \\
 82           & 127         & 209         &-17669       & 3831        &           &            & 596         &            &  \\
 82           & 128         & 210         &-14954       & 5355        &           &            &-428         & 9187        &  \\
 82           & 129         & 211         &-10522       & 3640        &           &            & 1379        & 8995        &  \\
 82           & 130         & 212         &-7602        & 5151        &           &            & 402         & 8791        &  \\
 82           & 131         & 213         &-2985        & 3453        &           &            & 2223        & 8605        &  \\
\end{tabular}
\begin{tabular}{llllllllll}
 86           & 126         & 212         &-8863        &           & 4403       &        &        &        & 7229 \\
 86           & 127         & 213         &-5816        & 5025        & 4498        &        &-2219        &        & 7842 \\
 86           & 128         & 214         &-4332        & 6586        & 5017        & 9454        &-3261        & 11611       & 8469 \\
 86           & 129         & 215         &-1216        & 4955        & 5138        & 8911        &-1517        & 11541       & 9134 \\
 86           & 130         & 216         & 378         & 6476        & 5673        & 8394        &-2500        & 11431       & 9800 \\
 86           & 131         & 217         & 3612        & 4837        & 5809        & 7846        &-731         & 11314       & 10492 \\
 86           & 132         & 218         & 5357        & 6325        & 6349        & 7331        &-1671        & 11163       & 11176 \\
 86           & 133         & 219         & 8744        & 4684        & 6490        & 6793        & 107         & 11010       & 11877 \\
\end{tabular}
\begin{tabular}{llllllllll}
 90           & 126         & 216         & 10652       &          & 2792       &        &        &         & 4027 \\
Z & N & A & $\Delta$M & S$_n$ & S$_p$ & Q$_\alpha$ & Q$_\beta$ &
S$_{2n}$ & S$_{2p}$
\\
\hline\\[-10pt]
 90           & 127         & 217         & 12483       & 6239        & 2879        &        &-4993        &         & 4626 \\
 90           & 128         & 218         & 12740       & 7814        & 3377        & 10212       &-6073        & 14054       & 5230 \\
 90           & 129         & 219         & 14567       & 6244        & 3483        & 9609        &-4410        & 14059       & 5863 \\
 90           & 130         & 220         & 14879       & 7759        & 3991        & 9061        &-5416        & 14003       & 6491 \\
 90           & 131         & 221         & 16787       & 6162        & 4107        & 8510        &-3713        & 13922       & 7140 \\
\end{tabular}
\begin{tabular}{llllllllll}
 94           & 126         & 220         & 36383       &           & 1243      &        &        &        & 950 \\
 94           & 127         & 221         & 37115       & 7339        & 1298        &        &-7591        &         & 1481 \\
 94           & 128         & 222         & 36267       & 8919        & 1756        & 11086       &-8718        & 16259       & 2014 \\
 94           & 129         & 223         & 36934       & 7403        & 1828        & 10490       &-7144        & 16323       & 2575 \\
 94           & 130         & 224         & 36096       & 8909        & 2296        & 9967        &-8185        & 16313       & 3131 \\
 94           & 131         & 225         & 36814       & 7353        & 2378        & 9457        &-6558        & 16263       & 3707 \\
 94           & 132         & 226         & 36086       & 8799        & 2850        & 9011        &-7530        & 16152       & 4274 \\
 94           & 133         & 227         & 36942       & 7214        & 2937        & 8574        &-5868        & 16014       & 4856 \\
\end{tabular}
\begin{tabular}{llllllllll}
 100          & 126         & 226         & 86124       &           &-973        &          &         &        &-3451 \\
 100          & 127         & 227         & 85484       & 8711        &-963         &         &-11109       &         &-3026 \\
 100          & 128         & 228         & 83260       & 10295       &-563         & 12740       &-12292       & 19006       &-2602 \\
 100          & 129         & 229         & 82476       & 8855        &-540         & 12171       &-10840       & 19150       &-2153 \\
 100          & 130         & 230         & 80195       & 10352       &-131         & 11690       &-11926       & 19207       &-1706 \\
 100          & 131         & 231         & 79407       & 8859        &-97          & 11236       &-10409       & 19211       &-1239 \\
 100          & 132         & 232         & 77193       & 10285       & 317         & 10857       &-11418       & 19144       &-777 \\
 100          & 133         & 233         & 76510       & 8754        & 359         & 10495       &-9853        & 19039       &-296 \\
\end{tabular}
\begin{tabular}{llllllllll}
 106          & 138         & 244         & 117797      & 10916       &-740         & 11388       &-12987       & 20572       &-2822 \\
 106          & 139         & 245         & 116489      & 9379        &-710         & 11239       &-11426       & 20296       &-2390 \\
 106          & 140         & 246         & 113949      & 10610       &-323         & 11127       &-12266       & 19990       &-1965 \\
 106          & 141         & 247         & 112951      & 9069        &-289         & 11005       &-10695       & 19680       &-1526 \\
 106          & 142         & 248         & 110746      & 10275       & 99          & 10915       &-11509       & 19345       &-1095 \\
 106          & 143         & 249         & 110082      & 8735        & 138         & 10810       &-9934        & 19011       &-650 \\
 106          & 144         & 250         & 108230      & 9923        & 529         & 10732       &-10727       & 18658       &-213 \\
 106          & 145         & 251         & 107915      & 8386        & 571         & 10635       &-9152        & 18310       & 235 \\
\end{tabular}
\begin{tabular}{llllllllll}
 114          & 154         & 268         & 164757      & 9753        &-663         & 13171       &-11629       & 18404       &-2488 \\
 114          & 155         & 269         & 164510      & 8318        &-629         & 13105       &-10166       & 18071       &-2081 \\
 114          & 156         & 270         & 163169      & 9412        &-268         & 13052       &-10896       & 17730       &-1679 \\
 114          & 157         & 271         & 163251      & 7988        &-229         & 12975       &-9437        & 17401       &-1265 \\
 114          & 158         & 272         & 162247      & 9075        & 134         & 12911       &-10157       & 17064       &-856 \\
 114          & 159         & 273         & 162655      & 7663        & 178         & 12822       &-8704        & 16738       &-434 \\
 114          & 160         & 274         & 161983      & 8743        & 545         & 12745       &-9413        & 16406       &-18 \\
 114          & 161         & 275         & 162712      & 7342        & 594         & 12644       &-7966        & 16086       & 409 \\
\end{tabular}
\renewcommand{\footnoterule}{\kern-3pt \hrule width .4\columnwidth
\kern 2.6pt}            
\end{table}

\end {document}